\documentclass[aps,prb,preprint,showpacs]{revtex4}
\usepackage{graphicx}
\usepackage{amssymb}
\usepackage{amsfonts}
\usepackage{amsmath}
\usepackage{dcolumn}
\usepackage{natbib}
\usepackage{color}

\def\vN{\vec{N}}

\begin{document}
\title{Machine Learning methods for interatomic potentials: application to boron carbide}
\author{Qin Gao$^1$, Sanxi Yao$^1$, Jeff Schneider$^2$, and Michael Widom$^1$}
\affiliation{Department of Physics$^1$, Robotics Institute$^2$, Carnegie Mellon University, Pittsburgh, Pennsylvania 15213, USA}

\begin{abstract}
Total energies of crystal structures can be calculated to high precision using quantum-based density functional theory (DFT) methods, but the calculations can be time consuming and scale badly with system size. Cluster expansions of total energy as a linear superposition of pair, triplet and higher interactions can efficiently approximate the total energies but are best suited to simple lattice structures.  To model the total energy of boron carbide, with a complex crystal structure, we explore the utility of machine learning methods ($L_1$-penalized regression, neural network, Gaussian process and support vector regression) that capture certain non-linear effects associated with many-body interactions despite requiring only pair frequencies as input. Our interaction models are combined with Monte Carlo simulations to evaluate the thermodynamics of chemical ordering.
\end{abstract}

\pacs{34.20.Cf,71.15.Nc,81.05.Zx}

\maketitle

\section{Introduction}
Density functional theory (DFT)~\cite{Hohenberg1964} can accurately determine the total energies of crystals. However, DFT is time consuming and scales as the cube of the number of atoms. It is thus generally infeasible to directly use DFT for energy prediction in the Monte Carlo (MC) simulation of phase transitions, which require many evaluations of the energy of very large structures. Interatomic potentials~\cite{Mihalkovic2012}, which typically fit the DFT energies as a function of the positions of the atomic nuclei, can be quickly evaluated to predict approximate energies. Cluster expansions~\cite{Sanchez84,Fontaine94}, which represent the energy as a sum of pair, triplet and higher-body interactions, provided a physically motivated and systematically improvable form for such a fit.  However, the number of necessary terms can grow quite rapidly for complex crystal structures with many inequivalent positions, so this approach has been most successful when applied to regular lattice structures. We notice that certain information in the higher order terms like triplets can be expressed as nonlinear functions of pairs. This motivates our study with machine learning (ML) methods such as regression models that can capture complex nonlinear interactions. ML methods have been used in various solid state physics problems~\cite{Curtarolo13,Seko14,Schutt14,Li15}. Interatomic potentials fitted with ML methods are generally more accurate and thus can be more useful for physical simulations~\cite{Lorenz04,Behler07,Bartok10,Rupp12}. 

\begin{figure}
\includegraphics[height=3in]{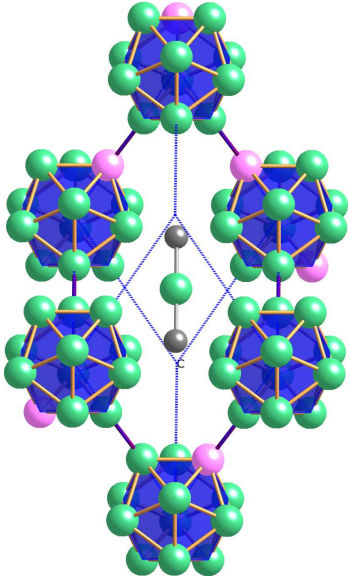}
\caption{\label{fig:BC}(Color online) Cut away view of boron carbide rhombohedral primitive cell showing 12-atom icosahedra and 3-atom chain. In the present model, all chains are C-B-C, while icosahedra have C randomly distributed among polar sites (shown in pink).}
\end{figure}

Boron carbide is an extremely hard and very light material with wide range of applications~\cite{Domnich11}.  Despite its importance, the phase diagram of boron carbide is not precisely known. Its complex structure (see Fig.~\ref{fig:BC}) features 12-atom icosahedra and 3-atom chains.  Because of their chemical similarity, substitutional disorder of boron and carbon is prevalent, in particular among the so-called {\em polar sites} of the icosahedra. Experimentally, boron carbide is difficult to equilibrate because of its strong covalent bonds. Additionally, the small difference between the atomic numbers of boron and carbon makes the precise composition and distribution of carbon atoms hard to measure. Two major problems exist in the widely accepted experimental boron carbide phase diagrams~\cite{Schwetz91,Okamoto92}: 1) the solubility range of carbon is given as $0.090 \le x_C \le 0.192$, while the DFT-predicted ground state has composition B$_4$C with $x_C=0.200$; 2) the boundaries of the composition range are temperature-independent, which is thermodynamically improbable.  Since experimental measurement is not reliable at low temperature (say, $T<$ 1000 K), computer simulation can help resolve these problems.

In this study, we exploit several ML methods to fit the interatomic potential of boron carbide and use the potential to perform MC simulations.  A previous study~\cite{Yao14}, fit a linear model of pair interactions for structures with $x_C=0.200$ ({\em i.e.} each icosahedron has a single polar carbon) and studied phase transitions at this high carbon limit. In this paper, instead of fixing $x_C=0.200$, we allow arbitrary substitutions and swaps between boron and carbon among the polar sites, thus extending the carbon concentration from a lower limit of B$_{13}$C$_2$ ($x_C=0.133$) with no upper limit. In particular, we allow {\em bipolar defects}~\cite{Vast09} in which two carbon atoms occupy a single icosahedron.  We fit the DFT energies with machine learning methods including $L_1$-penalized polynomial regression, neural network (NN), Gaussian process (GP) and support vector regression (SVR). We find the GP has smallest prediction error (0.31meV/atom), which is 33\% less than the linear model. We then perform Monte Carlo simulations with a linear regression model, a restricted polynomial regression model and a mixed-kernel Gaussian process model. The three energy models qualitatively agree with each other and indicate a phase transition at low temperature and high carbon chemical potential.

\section{Methods}

\subsection{DFT calculations}
Our calculation methods are similar to our previous study~\cite{Yao14}, namely electronic density functional theory utilizing PAW potentials~\cite{Blochl94,Kresse99} in the PBE~\cite{Perdew96} generalized gradient approximation with default energy cutoffs using VASP~\cite{Kresse93,Kresse96}. We calculate the fully relaxed total energies of 597 structures with supercell sizes $2\times2\times2$ (120 atoms) through $4\times4\times4$ (405 atoms).  The majority of the sampled structures were created through Monte Carlo simulations between $T=$400 and 2000K based on preliminary interaction models.  The data set to be fit consists of the enthalpies of formation of individual structures relative to the tie-line joining the ground states B$_{13}$C$_2$ and B$_4$C.

\subsection{Supervised Regression Models}
\subsubsection{Linear model}
One way of modeling the energies of structures is using cluster expansion~\cite{Sanchez84,Fontaine94,Paufler92,Walle02a}.  This approach is appealing because the relaxed energy is a function of the initial assignment of carbon atoms to polar sites, resulting in a lattice gas-type model. That is, only carbon positions must be specified, as boron necessarily occupy the remaining sites. However, due to the complexity of the boron carbide structure (a 15-atom basis) and the low density of polar carbons (typically 0-2 carbons among the 6 polar sites per cell), there are too many triplets and higher order clusters to be included.  A length cutoff of $R_c=6.58$~\AA~ would result in 427 triplets.  One feasible approximation is to truncate the expansion at the pairwise level, resulting in 23 pairs with separation up to $R_c$.  Our linear model (LM) based on this approximation is
\begin{equation}
E(\vN) = E(N_0,...,N_{23})=E_0+\sum_{i=0}^{23}{\beta_i N_i},
\label{eq:LM}
\end{equation}
where $N_0$ is the number of polar carbon in the structure and the remaining $N_i$'s are the different pairs.  The set $\{N_i\}$ can be considered as a 24-dimensional vector $\vN$. Appendix A1 presents some characteristics of the data set and its dependence on $\vN$.

\subsubsection{$L_1$-penalized polynomial model}
A polynomial model directly generalizes the linear model in Eq.~\ref{eq:LM}. Due to the limited size of the data set, we choose a second order polynomial model (PR2),
\begin{equation}
E(\vN) = E(N_0,...,N_{23})=E_0+\sum_{i=0}^{23}{\beta_iN_i}+\sum_{j=0}^{23}\sum_{k=j}^{23}{\gamma_{jk}N_jN_k},
\label{eq:PR2}
\end{equation}
which fully characterizes the second order interactions between numbers of pairs. This model contains 325 parameters. To avoid overfitting and perform feature selection, we add an $L_1$ norm penalty term. The resulting optimization problem is,
\begin{equation}
\min_{\vec{\theta}}\sum_{m=1}^{M}\frac{1}{2}(E^{DFT}_m - E(\vN_m;\vec{\theta}))^2+\lambda\|\vec{\theta}\|_1,
\label{eq:polyLasso}
\end{equation}
where $\vec{\theta}$ is the collection of all parameters, $E_0$, $\{\beta_i\}$ and $\{\gamma_{jk}\}$.  $E^{DFT}_m$ is the DFT calculated energy of the $m$th structure, $\vN_m$ is its 24 dimensional feature vector, $M$ is the size of the training set and  $\lambda$ is a tuning parameter. In this paper, we shall use $i$, $j$ and $k$ as indices for features, and $l$, $m$ and $n$ as indices for samples.

\subsubsection{Neural network}
In our neural network, the input layer contains 24 nodes corresponding to the components of $\vN$. We choose 1-2 hidden layers of 1-10 nodes with nonlinear activation functions like ``tanh" and ``sinh" in the hidden layers. The single node output layer utilizes a linear function. A Bayesian regularization for the model parameters reduces overfitting. We use the Matlab Neural Network Toolbox~\cite{Beale15} with default starting coefficients and regularization parameters. A detailed description of neural networks can be found in~\cite{Rasmussen06}.

\subsubsection{Gaussian process}
In GP, we assume the energies of structures are Gaussian distributed,
\begin{equation}
\left( \begin{array}{c}
E_{\rm train} \\
E_{\rm pred} \end{array} \right)\sim{\cal N} \left( \mu, \Sigma\right),\ {\rm with\ }\Sigma= \left(\begin{array}{cc}
\Sigma_{tt}\ \ \Sigma_{tp}\\
\Sigma_{tp}^T\ \ \Sigma_{pp}\end{array}\right),
\end{equation}
where the $E_{\rm train}$ and $E_{\rm pred}$ vectors denote the energies of training structures and structures whose energies to be predicted (predicting structures) respectively, $\cal N$ is a normal distribution with mean $\mu$ (set to zero in later derivation for simplicity), and $\Sigma$ is the covariance matrix. The mth row and nth column of $\Sigma$ is,
\begin{equation}
\Sigma_{mn} = k(\vN_m,\vN_n),
\end{equation}
where $\vN_m$ and $\vN_n$ are the feature vectors of the $m$th and $n$th structure respectively.  The kernel function $k(\vN_m,\vN_n)$ characterizes the similarity between feature vectors, and therefore structures. In our study, we use and compare the polynomial kernel $(1+\beta\vN_m\cdot\vN_n)^d$, the Gaussian kernel
${\rm exp}(-\|\vN_m-\vN_n\|_2^2/\gamma^2)$, and the Laplacian Kernel ${\rm exp}(-\|\vN_m-\vN_n\|_1/\gamma)$. 

A constant variance term $\delta^2$ is added to the kernel when $m=n$ to model the noise.  The DFT energy is precise and thus lacks noise. A perfect model containing all relevant inputs would in principle precisely fit the DFT energy. However, since our features only represent a subset of the structural information, the DFT energies are noisy in the subspace spanned by the features. We thus need variance $\delta^2$ to model the noise. The parameters $\beta$, $d$, $\gamma$ and $\delta$ in the kernels are called hyperparameters. We can optimize these hyperparameters by maximizing the likelihood of the training data, which is a convex optimization problem that can be efficiently solved~\cite{Rasmussen06}. We use the GPML Toolbox ~\cite{Rasmussen10} to perform our GP model fitting and testing.

Under the assumptions of GP, the conditional distribution of predicted energy given the training data is Gaussian.  We take the mean value of this distribution as the predicted energy. More explicitly, for a new structure with feature vector $\vN_l$, the predicted energy is,
\begin{equation}
E(\vN_l) = \sum_{m=1}^M\sum_{n=1}^M k(\vN_l,\vN_m)(\Sigma_{tt}^{-1})_{mn}E_{\rm train,n}.
\label{eq:GP}
\end{equation}
Moreover, GP also provides the variance of the predicted energy which implies the accuracy or confidence of the prediction, which is
\begin{equation}
\sigma^2(\vN_l) = k(\vN_l,\vN_l) - \sum_{m=1}^M\sum_{n=1}^M k(\vN_l,\vN_m)(\Sigma_{tt}^{-1})_{mn}k(\vN_n,\vN_l).
\end{equation}

\begin{figure}[ht]
\includegraphics[trim = 0mm 0mm 30mm 0mm, clip, scale=0.105]{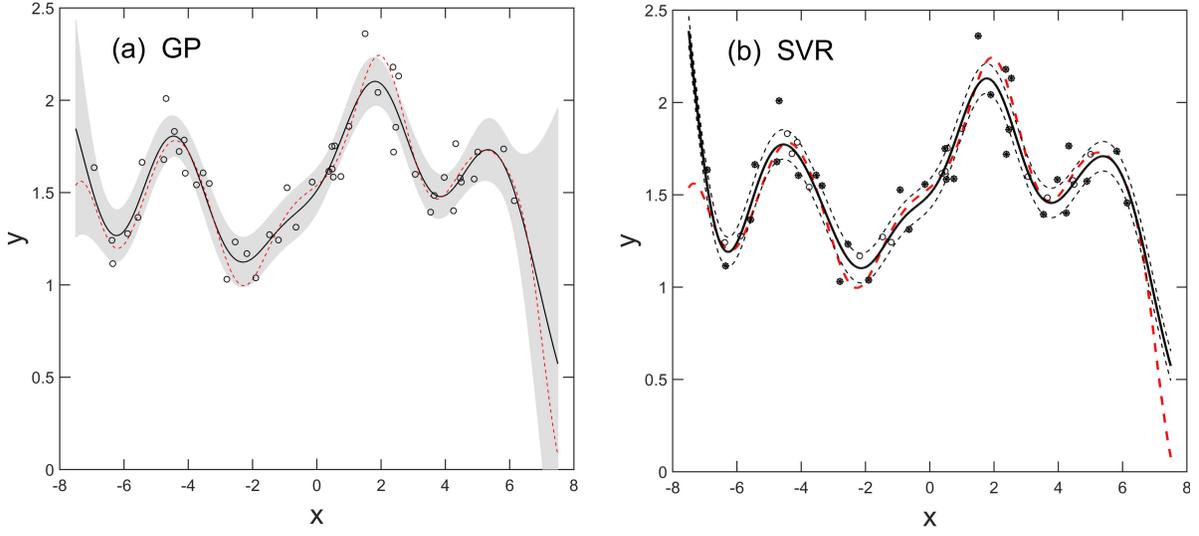}
\caption{ (a) GP and (b) SVR fits of a toy one-dimensional function. The dashed-red curve is the ground truth function. A ``o" denotes a data point which is drawn randomly from the ground truth function and added with Gaussian noise. The black solid lines in both panels are the fitted functions. (a) The shaded region lies within two standard error of the GP prediction (95\% confidence interval). (b) The dashed curves are the boundaries of the $\epsilon$-tube. The ``o" points with ``*" at the centers are support vectors.}
\label{fig:toyfunction}
\end{figure}

To illustrate the properties of GP, the left panel of Fig.~\ref{fig:toyfunction} shows a toy example of fitting a one dimensional function. The fitted nonlinear function captures the local properties of the data. Moreover, the fit provides the standard error of the prediction, where large standard error indicates small data density in the nearby region or large extrapolation. The standard errors of prediction can be used to check whether the data well sample the feature space, and to guide us in generating training structures for poorly represented regions.

\subsubsection{Support vector regression}

In SVR~\cite{Bishop07} the fitted function $E(\vN)=\sum_{i}\omega_i\cdot\Phi_i(\vN)+b$ minimizes the target function,
\begin{equation}
C\sum_{m=1}^M{\rm max}(0,|E(\vN_m)-E_m^{\rm DFT}|-\epsilon) + \frac{1}{2}\|\vec{\omega}\|_2^2,
\end{equation}
where $C$ and $\epsilon$ are positive real numbers, $\{\Phi_i(\vN)\}$ is a collection of chosen functions of $\vN$, and $\vec{\omega}$ and $b$ are fitting parameters. As shown in the right panel of Fig.~\ref{fig:toyfunction}, the errors of points within the $\epsilon$-tube are not counted in the target function, which is thus insensitive to the intrinsic small noise in the energies. The points outside of the $\epsilon$-tube only introduce linear penalty to the target function, which is more robust to outlier than least square error fit.

We can relax the constraints associated with the tube to obtain a dual form of SVR in which the energy prediction is
\begin{equation}
E(\vN_l)=\sum_{m=1}^M\alpha_mk(\vN_l,\vN_m)+b,
\label{eq:SVR}
\end{equation} 
where only data points on or outside the $\epsilon$ tube have nonzero $\alpha$ values, and are called support vectors. $b$ is a constant that can be calculated by support vectors.

One advantange of SVR is it transforms the features $\vec{N}$ to new features $\Phi_i(\vec{N})$, more suitable for fitting. However, $\{\Phi_i(\vec{N})\}$ might be a high (or even infinite) dimensional vector. That is hard to use in practice. In this dual form, instead of $\{\Phi_i(\vec{N})\}$, only the kernel function $k(\vN_l,\vN_m)=\sum_i \Phi_i(\vec{N}_l)\Phi_i(\vec{N}_m)$ is needed to train the model and make predictions. The possible kernel functions are similar as in GP, which are flexible and easy to use. We use the LIBSVM Toolbox ~\cite{Chang11} to perform our SVR model fitting and testing.

\subsection{Cross validation}
We perform 5-fold cross validation (CV) to evaluate our models. In 5-fold CV, the data is randomly divided into 5 sets. In every validation, we choose one set as the validation set and the rest as the training set. We train our model using the training set and predict the energies of the structures in the validation set. After such validations, we aquire the predicted energy of every structure in the dataset. We then calculate the CV root mean square errors (RMSE) of the structures, which is defined as,
\begin{equation}
RMSE = \sqrt{\frac{1}{M}\sum_{m=1}^M (E^{\rm DFT}_m - E^{\rm predict}_m)^2},
\end{equation}
where $M$ is the total number of structures in the dataset. CV RMSE mimics the generalization error, which is the standard error for predicting the energy of an unseen structure which is useful in comparing different models. Since ML models can be very complex, the fitting error can be very small while the CV error remains large, which is a sign of overfitting. Throughout this paper, we report the CV RMSEs and generalization errors, which are superior measures of model performance to fitting error.

\subsection{Monte Carlo}

Our Metropolis Monte Carlo simulations utilize a semi-grand canonical ensemble~\cite{Griffiths70,Kofke99,Widom12} in which we associate a chemical potential $\Delta\mu=\mu_C-\mu_B$ for the conversion of a polar boron to carbon.  The physically meaningful range of $\Delta\mu$ extends from -0.10 to +0.58 eV at $T=0$K based on enthalpies~\cite{Widom12}.  To circumvent difficulties arising from hysteresis near first order transitions, we apply replica exchange~\cite{Swendsen86} along both the temperature and chemical potential ($\Delta\mu/T$) axes.  Data was collected on grids in the $(T,\Delta\mu/T)$-plane in the form of multidimensional histograms $H(x_C,E;T,\Delta\mu/T)$ then was analyzed using multiple histogram methods~\cite{Ferrenberg89}.

\section{Results and analysis}

We calculate the RMSEs of 5-fold CV to evaluate the models as shown in Table~\ref{tab:CVerrors}.  Improvement is defined as the percentage decrease of RMSE compared to the linear model (LM, Eq.~(\ref{eq:LM})).  The second order polynomial regression with $L_1$ penalty (PR2, Eq.~(\ref{eq:PR2})) outperforms the linear model by a decrease of 20\% in RMSE error. The RMSE minimizes at 197 nonzero parameters out of a possible 325.  The neural network (NN) performs similarly to PR2 in CV. The best performing NN has 24 input nodes (features), one hidden layer of 3 or 4 nodes with ``tanh" activation function and a single-node output layer with linear activation function.  The nonparametric GP and SVR models (Eqs.~(\ref{eq:GP}) and~(\ref{eq:SVR})) decrease the CV error by around 33\%. Since GP has a probabilistic interpretation, we choose the hyperparameters by maximizing the likelihood of the training data. However, SVR does not have such interpretation, thus we perform an extensive search over a grid of hyperparameters to find the set of hyperparameters that minimizes the 5-fold CV error. 

\begin{table}[h!]
\begin{tabular}{| l | c | c | c|}
\hline
MODEL & RMSE (meV/atom) & improvement \\ \hline
LM & 0.48 $\pm$ 0.01  &0\%\\ \hline
PR2 & 0.39 $\pm$ 0.01 &20$\pm$1\% \\ \hline
NN & 0.37 $\pm$ 0.01 &24$\pm$1\%  \\ \hline
GP & \bf{0.31} $\pm$ 0.01 &\bf{33$\pm$1\%} \\ \hline
SVR & 0.32 $\pm$ 0.01 &32$\pm$1\%  \\ \hline
\end{tabular}
\caption{RMSE of 5-fold cross validation and improvement of different models. The standard error of these quantities are obtained from the statistics of ten repetitions of cross validation.}
\label{tab:CVerrors}
\end{table}

\begin{figure}[ht]
\includegraphics[trim = 0mm 0mm 0mm 0mm, clip, scale=0.35]{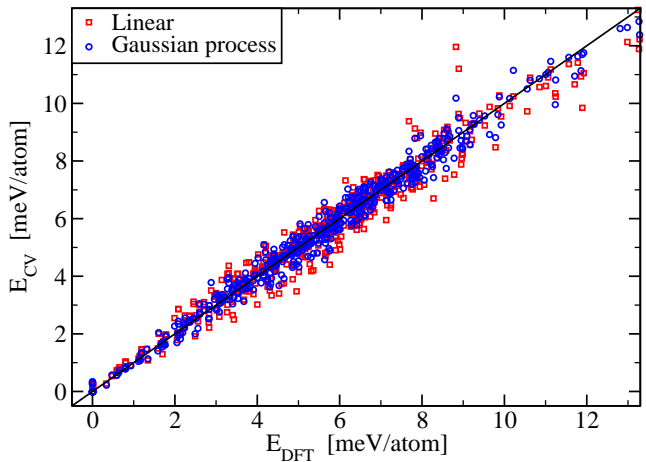}
\caption{(Color online) The predicted energies in 5-fold CV vs the ground truth DFT energies. The black line is $y=x$. Red points are from linear model and blue from GP predictions.}
\label{fig:goodnessoffit}
\end{figure}

The goodness of fit is shown in Fig.~\ref{fig:goodnessoffit} by comparing the predicted energies of the validation sets in the 5-fold CV with the corresponding DFT energies. The points generally lie near the $y=x$ line. GP fits better than the linear model. To illustrate the fine details, we only show structures with energies up to 13.3 meV/atom rather than the maximum energy of around 40 meV/atom present in the full data set.  For comparison, the mean energy at a high temperature of $T=2500$K is around 10 meV/atom in the high carbon limit.  To further compare the performance of linear model and GP, the residuals of the validation sets in the 5-fold CV are shown in Appendix A2. The residuals of the linear model are generally larger than residuals of GP.  Patterns exist in the residuals which indicate underfitting, and these are more obvious in the linear model.

Since we need to predict energies of large cell structures in our Monte Carlo simulation, we also study the performance of our models when generalizing to large cells. We use our $2\times2\times2$ and $3\times3\times3$ cell structures as the training set and the remaining 12 larger cell structures ($3\times3\times4$ and $4\times4\times4$) as the generalization set. The generalization error are 0.43, 0.46 and 0.86 meV/atom for SVR, GP and the linear model respectively.  All these models have larger generalization errors than the CV errors of the whole dataset, but this difference is less pronounced for SVR and GP.

\subsection{Model Selection and Acceleration}
\begin{figure}[ht]
\includegraphics[trim = 0mm 0mm 0mm 0mm, clip, scale=0.45]{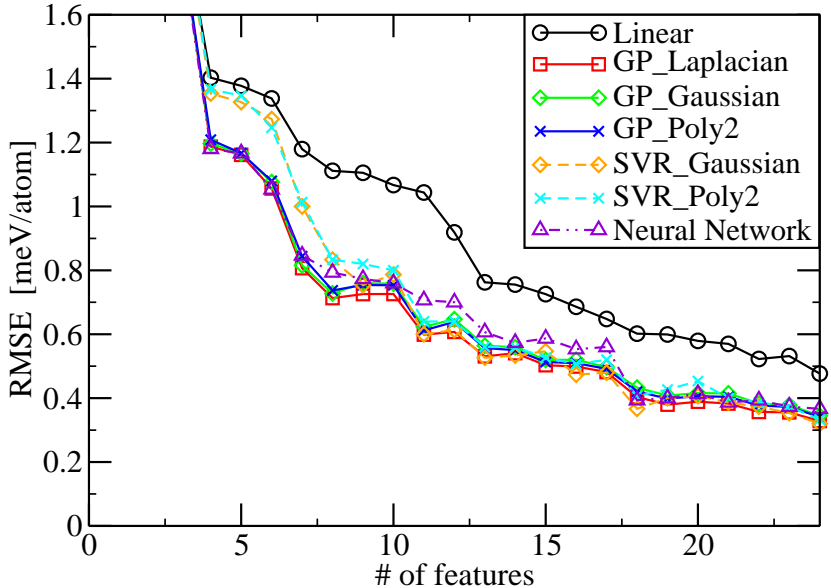}
\caption{(Color online) RMSE of linear model, GP, NN and SVR vs number of features. The black curve is linear model, solid curves are GP with different kernels. Dashed curves are SVR with different kernels. The purple dot-dashed curve is NN.}
\label{fig:RMSEvsNfeature}
\end{figure}

We perform a greedy stepwise feature selection, as shown in Fig.~\ref{fig:RMSEvsNfeature} where the 24 features yield the smallest CV errors. Note that the CV error is still decreasing near 24 features, which suggests that our description of the structures with these 24 features is insufficient and further improvement could be made by adding more effective features. Moreover, Fig.~\ref{fig:RMSEvsNfeature} shows that both GP and SVR CV errors are insensitive to the choice of kernel. 

Since the CV error is insensitive to the choice of kernel, we use GP with polynomial kernel of degree $d=2$ to predict the energies quickly enough for the Monte Carlo themodynamics simulation. In Metropolis Monte Carlo simulation, structures are generated and energies are claculated sequentially and energies.  Directly using Eq.~\ref{eq:GP} or Eq.~\ref{eq:SVR} to predict is slow since to predict one energy we have to sum over the whole training set ($\sim$ 600) or all support vectors ($\sim$ 400). It thus takes several hundreds of inner products of 24-dimensional vectors to predict one energy. However, with a polynomial kernel of degree two, we can use a change of summation order trick to accelerate the prediction, which is essentially rewriting the prediction in a parametric form. 

Defining $\vec{\alpha} = \Sigma_{tt}^{-1}E_{\rm train}$, which can be easily calculated offline before the Monte Carlo simulations, the GP energy prediction Eq.~(\ref{eq:GP}) can be rewritten as
\begin{equation}
E(\vN_l) = \sum_{m=1}^M \alpha_m k(\vN_l,\vN_m)= \sum_{m=1}^M \alpha_m(1+\beta\vN_l\cdot\vN_m)^2 = c + \vec{v}\cdot\vN_l + \vN_l^T A\vN_l,
\label{eqn:GPMC}
\end{equation}
where $c=\sum_{m=1}^M \alpha_m$, $\vec{v} = \sum_{m=1}^M 2\beta\alpha_m\vN_m$ and the matrix $A=\sum_{m=1}^M\beta^2 \alpha_m\vN_m\vN_m^T$. Since $c$, $\vec{v}$ and $A$ can be calculated in advance using the training set, the calculation to predict one structure only needs 25 vector multiplications, which is 30 times fewer than directly using Eq.~\ref{eq:GP}. In practice, the prediction is fast enough for Monte Carlo simulation.  The same trick works for SVR also.

\section{Monte Carlo simulation}

During MC simulation, we find that the GP model in Eq.~\ref{eqn:GPMC} predicts unphysically low energies for some structures. This is because the dataset size is limited and the distributions of the numbers of long bonds (e.g. the values of $N_i$ for large $i$) is large, leading to unreliable extrapolation.  We found two solutions to this problem.

First, we modify our GP model by defining a mixed kernel that captures interactions between carbon concentration ($N_0$) and the three types of shortest bonds, while ignoring other interactions.  Defining $\vN_{\parallel}=(N_0,N_1,N_2,N_3)$ and $\vN_{\perp}=(N_4,N_5,...,N_{23})$, our mixed Gaussian/linear kernel is
\begin{equation}
k(\vN,\vN') = \sigma_f^2 {\rm exp}(\frac{||\vN_{\parallel}-\vN'_{\parallel}||_2^2}{2l^2})+\sigma_p^2(\vN_{\perp}\cdot\vN'_{\perp}+c),
\end{equation}
where $\sigma_f$, $l$, $\sigma_p$ and $c$ are hyperparameters. This kernel has RMSE 0.36$\pm$ 0.01 meV/atom in 5-fold CV which is slightly higher than the 0.31$\pm$0.01 meV/atom obtained from GP with full Gaussian kernel.

Second, since we observed that the GP model with polynomial kernel of degree $d=2$ is similar to a parametric polynomial regression of degree two (PR2), we fit less flexible PR2 models with only short-bond interactions. We start with the 24 features $N_i$ and add the products $N_iN_j$ successively as new features for $i = 0, 1, 2$ and $j = i, i+1,  ..., 10$. The 41-feature PR2 model has smallest CV error, which is 0.35$\pm$0.01 meV/atom, similar to the mixed-kernel GP. 

\begin{figure}[ht]
\includegraphics[scale=0.33]{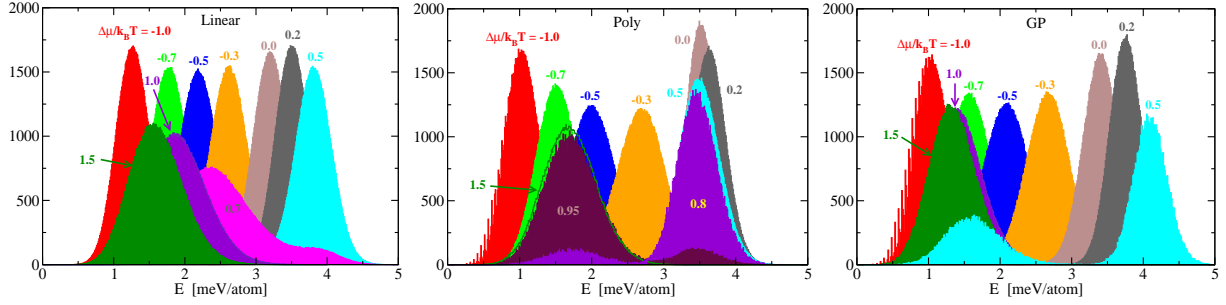}
\caption{(Color online) Energy histograms of the linear model (left), the 41-feature PR2 model (middle), and the mixed-kernel GP (right).}
\label{fig:hist}
\end{figure}

Simulations of a variety of supercell sizes were performed using energies predicted by the linear model, the mixed-kernel GP, and the 41-feature PR2 model.  Resulting histograms of energies of 6$\times$6$\times$6 cells at $T = 600 K$ and different $\Delta\mu/k_BT$'s are shown in Fig.~\ref{fig:hist}. Rapidly changing histograms, or multiply-peaked histograms indicate a possible phase transition. The histograms of these models are similar in trend, with rapid change between $\Delta\mu/k_BT$=0.5 and 1.0.

\begin{figure}[ht]
\includegraphics[scale=0.07]{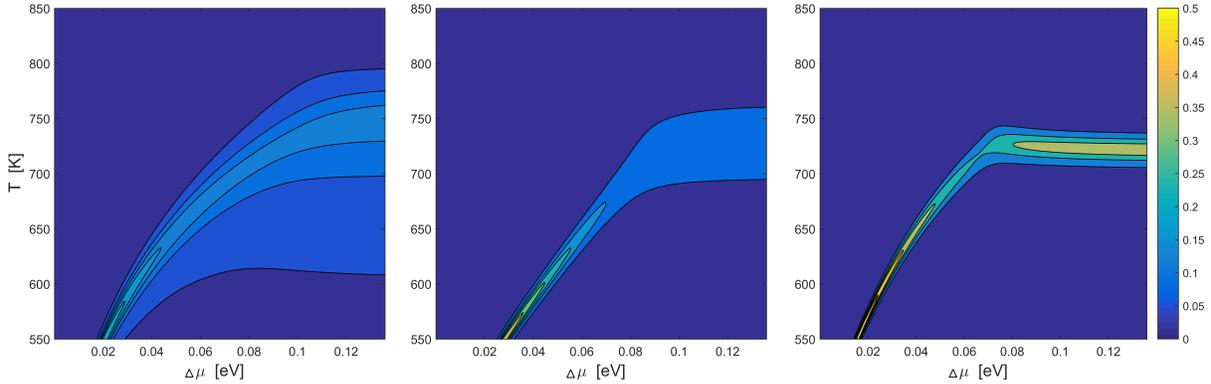}
\caption{(Color online) MC simulated heat capacity of boron carbide using the linear model (left), the 41-feature PR2 model (middle), and the mixed-kernel GP (right).}
\label{fig:HeatCapacity}
\end{figure}

Based on the histograms in the MC simulations, the heat capacities of boron carbide are calculated using the multi-histogram method~\cite{Ferrenberg88,Ferrenberg89,Yao14}, as shown in Fig.~\ref{fig:HeatCapacity}. At infinite cell size, diverging heat capacity indicates a phase transition. We analyzed the scaling of heat capacity with 4$\times$4$\times$4, 5$\times$5$\times$5 and 6$\times$6$\times$6 cells. The peaks of the heat capacities in all three models increase quickly and do not greatly alter their positions. All these models thus suggest a first order phase transition, which corresponds well with the fast evolving histograms in high $\Delta\mu/{k_B}T$, seen in Fig.~\ref{fig:hist}. This phase transition is consistent with a previous study~\cite{Yao14} at 20\% carbon limit. The mixed-kernel GP, and the 41-feature PR2 model more accurately predict the energies, thus should be quantitatively more reliable. The previously found second order transition~\cite{Yao14} is hard to see using heat capacity. 

\begin{figure}[h]
\includegraphics[scale=0.05]{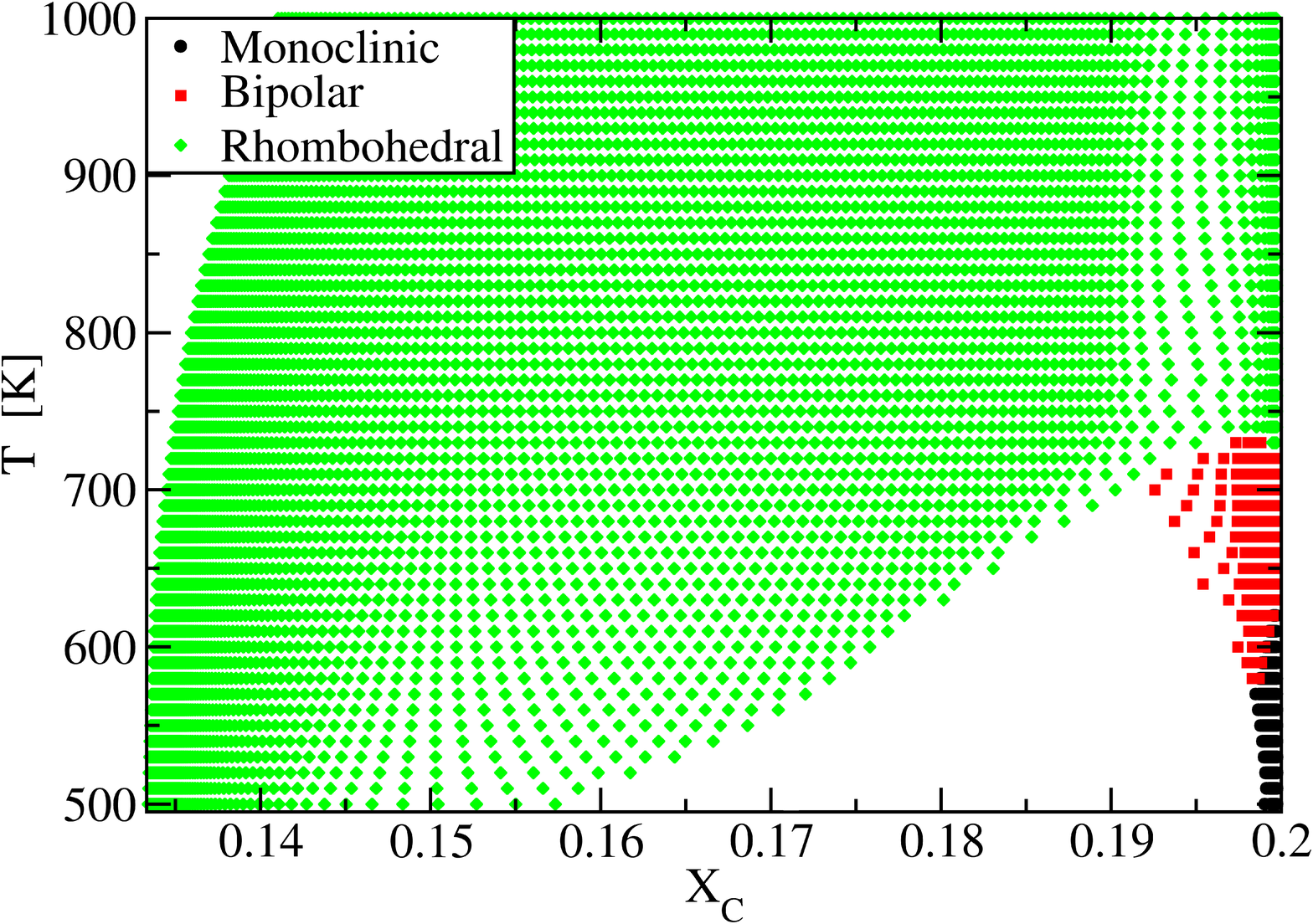}
\caption{(Color online) Boron carbide phase diagram in the ($X_C$, T) plane. Black region is the monoclinic phase, red is the bipolar phase, and green is the rhombohedral phase.}
\label{fig:BCphase}
\end{figure}

Fig.~\ref{fig:BCphase} shows the phase diagram of boron carbide obtained from our MC simulation with the mixed-kernel GP interaction model. Three phases appear~\cite{Yao14}: 1) Rhombohedral phase, where carbon occupies the six polar sites on the icosahedra with equal probability; 2) Bipolar phase, where the 3-fold rotational symmetry is broken but the two poles remain equivalent; 3) Monoclinic phase, where the 2-fold symmetry between the two poles is broken in addition to the 3-fold rotational symmetry. At lower temperature, the Rhombohedral phase (green) shrinks gradually to the point $X_C=0.133$ as $T\rightarrow 0$ K, and the Monoclinic phase (black) shrinks gradually to the point $X_C=0.20$ as $T\rightarrow 0$ K. The rhombohedral phase extends down to $X_C=0.133$ at all temperatures. The existence of three phases agrees with a previous restricted study at 20\% carbon composition~\cite{Yao14}, although here we find a bipolar phase where the restricted model instead exhibited a polar phase owing to its exclusion of bipolar defects.  A comprehensive study of the phase transitions and analysis of the order parameters will be reported in a forthcoming paper. 

\section{Conclusion}

In this study we construct ML-based interatomic potentials of boron carbide and perform MC simulations with these potentials. We started with a cluster-expansion motivated linear model, then exploit the nonlinear interaction between features using parametric models like $L_1$-regularized polynomial model and neural network, and nonparametric models like Gaussian process and support vector regression. Our result shows that $L_1$-regularized polynomial model and neural network decrease the cross validation error by 20\%, while the nonparametric models achieve 33\% improvement. The accuracy of nonparametric models is insensitive to the choice of kernel in our problem.

We perform MC simulation with gaussian process (GP). Directly using polynomial-kernel GP leads to wrong ground states.  Augmenting our data set to include structures with large predicted uncertainty resulted in decreased CV RMSE but did not alleviate the problem of false ground states. Instead we developed a mixed-kernel GP model and a restricted polynomial regression model to fix it. Our MC simulations with these models and the linear model all clearly indicate a phase transition at high carbon concentration and low temperature. To improve the models, some local properties of atoms could be included in the models, rather than just using the total numbers of pairs in the structure. Another possible improvement is to include numbers of triplets. Including such local features or triplets might require more data to avoid extrapolation problems.

\section{Acknowledgments}
Financial support from the McWilliams Fellowship at CMU and from ONR-MURI under the grant NO. N00014-11-1-0678 is gratefully acknowledged.

\bibliographystyle{apsrev}
\bibliography{BC}

\section{Appendix}

\subsection{A1. Data Exploration}

\begin{figure}[ht]
\includegraphics[trim = 0mm 0mm 0mm 0mm, clip, width=16.5cm]{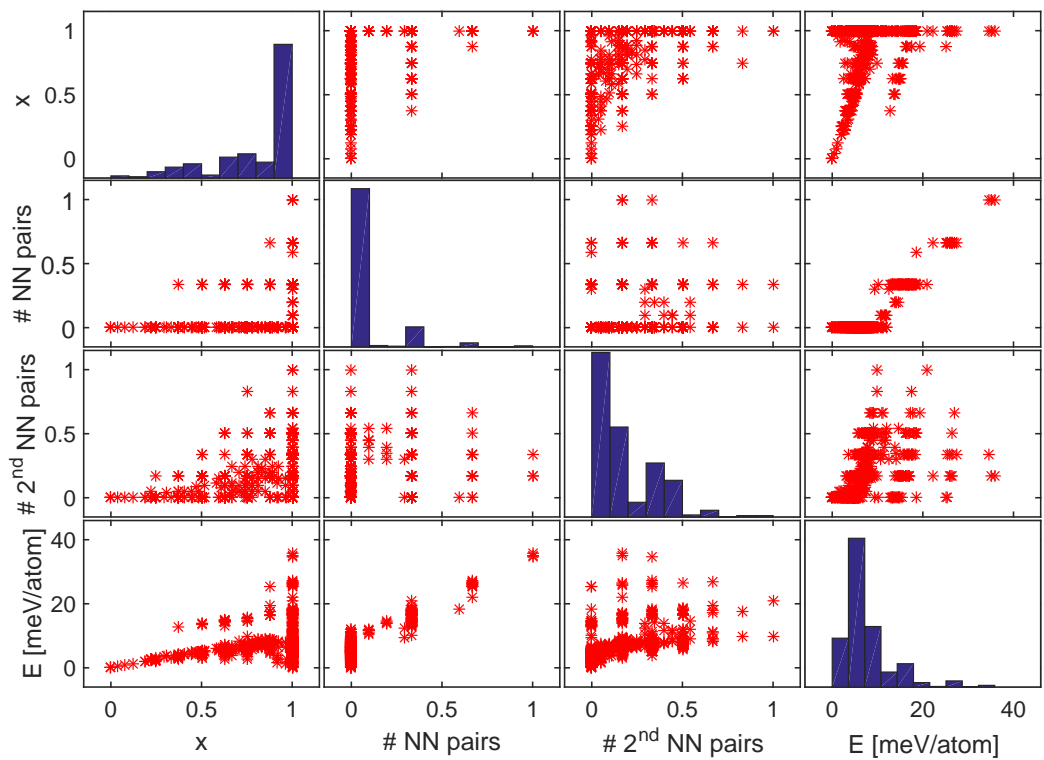}
\caption{(Color online) Histograms (diagonal plots) and pairwise scatter plots (off-diagonal plots) of features and energies. The variables from left to right and from top to bottom are the polar C concentration $x_C$, number of nearest C-C bonds $N_1$, number of second nearest C-C bond $N_2$, and the DFT calculated energy $E$. All the three features are normalized. Every red star represents the value of one structure.}
\label{fig:EDA}
\end{figure}

Histograms and pairwise scatter plots of three selected features (polar C concentration $x$, number of nearest C-C bonds $N_1$, number of second nearest C-C bonds $N_2$) and energy $E$ are shown in Fig.~\ref{fig:EDA}. All the variables are skew distributed due to the physical distribution of the boron carbide structures. For example, our structrues are weighted towards few short bonds $N_1$ and low energy E. Moreover, the scatter plots show the features are correlated but not collinear and that the energy correlates with these three features. For example, low energy correlates with low $N_1$. The variances of energies at different feature values are not the same. The scewed distributions and non-constant variances might bring difficulty to linear model and other models with similar assumptions.

\subsection{A2. Residuals of Linear model and GP}
\begin{figure}[ht]
\includegraphics[trim = 40mm 0mm 0mm 0mm, clip,scale=0.4]{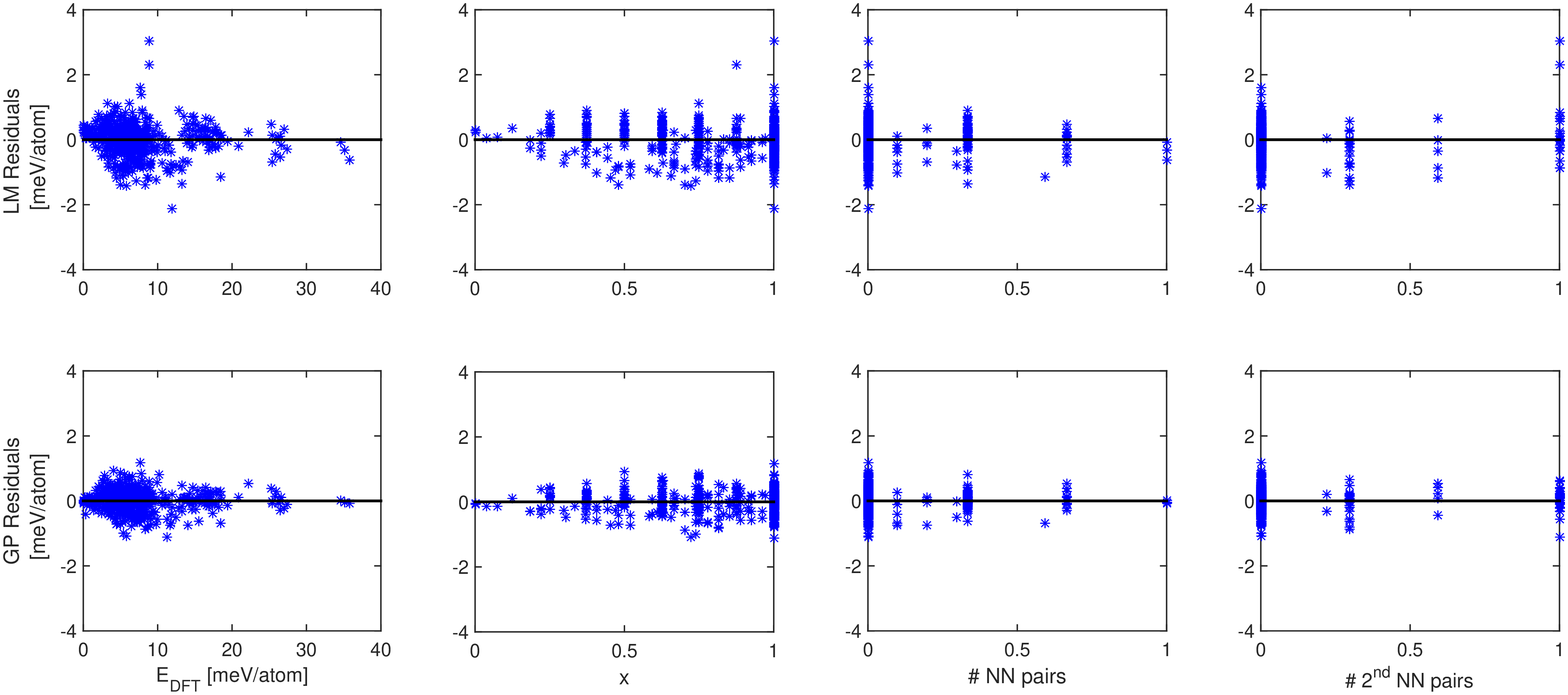}
\caption{(Color online) Residuals of predicted energies in 5-fold CV of linear model (top) and GP (bottom). From left to right the x-axis represents the DFT energy, carbon concentration, number of nearest bonds and number of second nearest bonds, respectively.}
\label{fig:resid}
\end{figure}

The residuals of the linear model and the GP are compared in Fig.~\ref{fig:resid}. The residuals of GP are smaller in magnitude, and less pronounced patterns exist in GP than in the linear model.

\end{document}